\documentclass[aps,prb,twocolumn,superscriptaddress,showkeys,amsmath,amssymb]{revtex4-1}
\usepackage{graphicx}
\usepackage{upgreek,mathtools}
\usepackage{booktabs}
\usepackage{color}
\usepackage{hyperref}

\usepackage{makecell}
\usepackage{siunitx}
\makeatletter
\newcommand\footnoteref[1]{\protected@xdef\@thefnmark{\ref{#1}}\@footnotemark}
\makeatother

\begin{document} 

\title[E-Ph Coupling and Surface Debye Temperature of Bi$_2$Te$_3$(111)]{Electron-Phonon Coupling and Surface Debye Temperature \\ of Bi$_2$Te$_3$(111) from Helium Atom Scattering}

\author{Anton Tamt\"{o}gl}
\email{tamtoegl@gmail.com}
\affiliation{Cavendish Laboratory, J. J. Thompson Avenue, Cambridge CB3 0HE, United Kingdom}
\affiliation{Institute of Experimental Physics, Graz University of Technology, 8010 Graz, Austria}
\author{Patrick Kraus}
\affiliation{Institute of Experimental Physics, Graz University of Technology, 8010 Graz, Austria}
\author{Nadav Avidor}
\affiliation{Cavendish Laboratory, J. J. Thompson Avenue, Cambridge CB3 0HE, United Kingdom}
\author{Martin Bremholm}
\author{Ellen M. J. Hedegaard}
\author{Bo B. Iversen}
\affiliation{Center for Materials Crystallography, Department of Chemistry and iNANO, Aarhus University, 8000 Aarhus, Denmark}
\author{Marco Bianchi}
\author{Philip Hofmann}
\affiliation{Department of Physics and Astronomy, Interdisciplinary Nanoscience Center (iNANO), Aarhus University, 8000 Aarhus, Denmark}
\author{John Ellis}
\author{William Allison}
\affiliation{Cavendish Laboratory, J. J. Thompson Avenue, Cambridge CB3 0HE, United Kingdom}
\author{Giorgio Benedek}
\affiliation{Dipartimento di Scienza dei Materiali, Universit\`{a} degli Studi di Milano-Bicocca, 20125 Milano, Italy}
\affiliation{Donostia international Physics Center (DIPC), University of the Basque Country (EHU-UPV), 20018 Donostia - San Sebastian, Spain}
\author{Wolfgang E. Ernst}
\affiliation{Institute of Experimental Physics, Graz University of Technology, 8010 Graz, Austria}

\begin{abstract}
We have studied the topological insulator Bi$_2$Te$_3$(111) by means of helium atom scattering. The average electron-phonon coupling $\lambda$ of Bi$_2$Te$_3$(111) is determined by adapting a recently developed quantum-theoretical derivation of the helium scattering probabilities to the case of degenerate semiconductors. Based on the Debye-Waller attenuation of the elastic diffraction peaks of Bi$_2$Te$_3$(111), measured at surface temperatures between $110~\mbox{K}$ and $355~\mbox{K}$, we find $\lambda$ to be in the range of $0.04-0.11$. This method allows to extract a correctly averaged $\lambda$ and to address the discrepancy between previous studies. The relatively modest value of $\lambda$ is not surprising even though some individual phonons may provide a larger electron-phonon interaction.\\
Furthermore, the surface Debye temperature of Bi$_2$Te$_3$(111) is determined as $\Uptheta_D = (81\pm6)~\mbox{K}$. The electronic surface corrugation was analysed based on close-coupling calculations. By using a corrugated Morse potential a peak-to-peak corrugation of 9\% of the lattice constant is obtained.
\end{abstract}

\keywords{Bi2Te3, Topological Insulator, Electron-phonon coupling, Debye temperature, Helium atom scattering}

\pacs{63.20.kd,68.49.Bc,68.35.Ja,73.20.-r}

\maketitle 

\section{Introduction}
Bi$_2$Te$_3$ is a layered narrow-gap ($\approx 0.2~\mathrm{eV}$) semiconductor which is widely studied as a thermoelectric material\cite{Goldsmid2014}. The excellent thermoelectric performance of Bi$_2$Te$_3$ has been attributed to the details of the electronic structure and a low lattice thermal conductivity similar to ordinary glass. Many recent advances in enhancing the thermoelectric figure of merit are linked with nanoscale phenomena found in bulk samples with nanoscale constituents as well as in nanoscale samples and thin film devices\cite{Snyder2008,Pettes2013}.\\ 
More recently, Bi$_2$Te$_3$ has been classified as a topological insulator\cite{Chen2009}, a recently discovered class of materials with an insulating bulk electronic structure and protected metallic surface states\cite{Hasan2010,Moore2009}. Surface-dominated transport is a major objective on the way to technical applications of these materials\cite{Barreto2014}, however, the detailed scattering channels for the surface state electrons may impose constraints for potential applications. In particular, the electron-phonon (e-ph) interaction may introduce a strong scattering mechanism at finite temperatures\cite{Huang2012,Chen2013,Sobota2014}. The e-ph coupling constant $\lambda$ is a convenient and effective parameter to characterise the strength of interaction between electrons and phonons on these surfaces.\\
While the e-ph coupling on topological insulator surfaces has been widely studied recently\cite{Zhu2012,Hatch2011,Pan2012,Kondo2013,Sobota2014}, conflicting values of $\lambda$ have been reported for the surface of Bi$_2$Te$_3$(111)\cite{Kondo2013,Howard2013,Chen2013}. The large variability of the mass-enhancement factor $\lambda$ from different experimental and theoretical sources depends on the balance of the two kinds of charge carriers, i.e. on the actual position of the Fermi level: If the Fermi level falls into the gap (ordinary semiconductor), surface Dirac states dominate whereas for a degenerate semiconductor the surface conduction electrons dominate. Even when the Fermi level falls into the gap, the doping for a specific surface can be expected to play an important role as well: While the phase space for scattering can be quite large for a strongly doped sample, it shrinks to nearly zero when the doping is such that the Dirac point is placed at the Fermi level.\\
We attempt to clarify this discrepancy among the experimental results by applying a new approach which allows to extract the correctly averaged $\lambda$\cite{Manson2016}. Our approach is based on helium atom scattering (HAS) from Bi$_2$Te$_3$(111), where in the degenerate n-type semiconductor case, the contribution of the surface two-dimensional electron gas (2DEG) overwhelms that of the Dirac states.\\
He atoms at thermal energies are scattered directly by the surface charge density corrugation rather than the ion cores. In the case of inelastic scattering, this corresponds to a scattering by phonon-induced charge density oscillations. This means in turn that the measured intensity of a phonon mode can be used to infer information about the e-ph coupling strength for a specific phonon\cite{Tamtogl2013a,Benedek2014}, resulting in a mode-specific $\lambda$. Recently Manson \textit{et al.} \cite{Manson2016} showed that the Debye-Waller (DW) exponent in atom scattering from a conducting surface can be directly related to the e-ph coupling constant $\lambda$. Since the DW exponent contains the sum over all contributing phonon modes, this theory can be used to determine the average e-ph coupling $\lambda$.\\
We use this approach, originally formulated for a 3D electron gas model, in a new formulation adapted to the surface of a degenerate semiconductor to extract the e-ph coupling strength from temperature-dependent He atom scattering measurements. Furthermore, we present a study of the surface Debye temperature of Bi$_2$Te$_3$(111). Finally, diffraction peak intensities are used to extract the surface electronic corrugation upon scattering of a thermal energy He beam.

\section{Experimental Details}
All measurements were performed on the Cambridge helium-3 spin-echo apparatus which generates a nearly monochromatic beam of $^3$He that is scattered off the sample surface in a fixed 44.4$^{\circ}$ source-target-detector geometry. The detailed setup of the apparatus has been described in greater detail elsewhere\cite{Alexandrowicz2007,Jardine2009}. The reported measurements were carried out using an incident beam energy of 8 meV.\\
The crystal structure of Bi$_2$Te$_3$ is rhombohedral, consisting of quintuple layers (QL) bound to each other through weak van der Waals forces which gives easy access to the (111) surface by cleavage\cite{Michiardi2014}. The (111) cleavage plane is terminated by Te atoms and exhibits a hexagonal structure ($a=4.386~\mathrm{\AA}$)\cite{Huang2012}. The Bi$_2$Te$_3$ single crystals used in the study were attached onto a sample holder using electrically and thermally conductive epoxy. The sample holder was then inserted into the chamber using a load-lock system\cite{Tamtogl2016a} and cleaved \textit{in-situ}. The sample holder can be heated using a radiative heating filament on the backside of the crystal or cooled down to $100~\mbox{K}$ using liquid nitrogen. The sample temperature was measured using a chromel-alumel thermocouple.\\
Angle-resolved photoemission spectroscopy (ARPES) data were taken at the SGM-3 beamline of the synchrotron radiation facility ASTRID2 in Aarhus \cite{Hoffmann2004}. The combined energy resolution of beamline and analyser was better than 30~meV, and the experimentally determined angular resolution was better than 0.2$^{\circ}$. The sample temperature was 100~K for all ARPES data shown here.

\section{Results and discussion}

\subsection{Helium diffraction from Bi$_2$Te$_3$(111)}
\label{sec:elastic}
\begin{figure}
\centering
\includegraphics[width=0.48\textwidth]{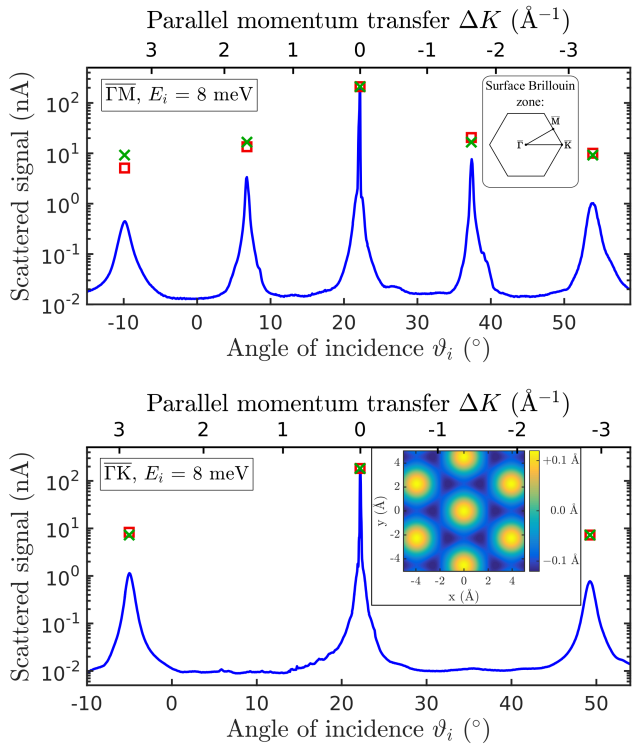}
\caption{Scattered He intensities (logarithmic scale) for Bi$_2$Te$_3$(111) versus incident angle $\vartheta_{i}$ for the two high symmetry directions. The upper panel shows a scan along the $\overline{\mathrm{\Gamma M}}$ azimuth and the lower panel a scan along  $\overline{\mathrm{\Gamma K}}$. The upper abscissa shows the corresponding parallel momentum transfer. In both cases the crystal was kept at 110 K and an incident beam energy of 8 meV was used. The inset in the upper panel shows the surface Brillouin zone and the scanning directions.\\
The red squares show the peak areas determined from measurements (scaled relative to the specular intensity). They are in good agreement with best-fit results from close-coupling calculations (green crosses), except for a slight asymmetry in the angular scan along $\overline{\mathrm{\Gamma M}}$. The inset in the lower panel shows a contour plot of the surface corrugation as obtained by the best-fit results from the close-coupling calculation.}
\label{fig:Diffraction}
\end{figure}

\autoref{fig:Diffraction} displays the scattered He intensity versus the incident angle $\vartheta_{i}$ on Bi$_2$Te$_3$(111) for both high symmetry directions of the crystal. The crystal was held at 110 K for both measurements and the intensity is shown on a logarithmic scale. The lattice constant as determined from the diffraction peak positions in \autoref{fig:Diffraction} is $a = ( 4.36 \pm 0.02 )~\mbox{\AA}$. This is in good agreement with the reported values in literature  ($a=4.386~\mathrm{\AA}$  at room temperature \cite{Huang2012}) and has also been used for the calculations presented in \autoref{sec:ElCorrug}. The specular peak ($\vartheta_i = \vartheta_f = 22.2^{\circ}$) in \autoref{fig:Diffraction} exhibits a full width at half maximum FWHM $=(0.095\pm0.005)^{\circ}$ which corresponds to a width of $0.02~\mathrm{\AA}^{-1}$ in terms of the momentum transfer. \\
In general the width of the specular peak is determined by the angular broadening of the apparatus and the quality of the crystal\cite{Becker2010,Comsa1979,Lapujoulade1980}. Hence a measurement of the angular spread in the specular peak provides an estimate of the surface quality. This is possible since the peak broadening is proportional to the average domain size, also known as the surface coherence length. The measured specular width $\Delta \theta_{exp}$ is a convolution of the angular broadening of the apparatus $\Delta \theta_{app}$ and the domain size broadening $\Delta \theta_{w}$ given by:  $\Delta \theta_{exp}^2  =\Delta \theta_{w}^2+ \Delta \theta_{app}^2$. The coherence length can then be determined using:
\begin{equation}
l_c = \frac{5.54}{\Delta \theta_w \, k_i \, \cos \vartheta_f }
\label{eq:CohLength}
\end{equation}
with $k_i$ the wavevector of the incoming He beam and $\vartheta_f$ the final scattering angle\cite{Becker2010,Comsa1979,Lapujoulade1980}. The angular broadening $\Delta \theta_{app}$ of the He spin-echo apparatus is almost the same size as the experimental broadening. Nevertheless, we can use the experimentally determined broadening $\Delta \theta_{exp}$ together with \eqref{eq:CohLength} to obtain an estimate (lower limit) for the quality of the crystal. The measured FWHM of the specular peak gives rise to an average domain size of at least $1000~\mathrm{\AA}$. This illustrates that topological insulator surfaces can be prepared with an exceptionally good surface quality by in-situ cleaving and are perfectly suited for atom scattering studies. Compared to previous HAS experiments\cite{Howard2013} the FWHM of the specular peak is much smaller (by a factor of approximately 20) which indicates the high quality of the investigated surface.\\

\subsection{Electronic Corrugation}
\label{sec:ElCorrug}
Since He atoms are scattered by the charge distribution at the surface, the shape of this distribution influences the height of the diffraction peaks. The surface charge density as ``seen'' by He atoms at a certain incident energy can be written as $\xi(\mathbf{R})$ where $\mathbf{R}$ is the lateral position in the surface plane. $\xi(\mathbf{R})$ is referred to as the electronic surface corrugation as it describes a periodically modulated surface with constant total electron density (see the inset in the lower panel of \autoref{fig:Diffraction}).\\
The electronic corrugation is usually obtained by calculating the expected diffraction probabilities for a certain $\xi(\mathbf{R})$ and then optimising $\xi(\mathbf{R})$ to get the best match between the theoretically and experimentally obtained peak heights.\\
The elastic diffraction peaks of He scattered from Bi$_2$Te$_3$(111) were calculated using the exact close-coupling (CC) method\cite{Sanz2007} which goes beyond the simple hard-wall approximation and has been mainly applied to ionic and metallic surfaces\cite{Farias1998}. 
An exact description of the CC formalism can be found in the literature\cite{Sanz2007} and its application to a hexagonal semimetal surface is described in Mayrhofer-Reinhartshuber \textit{et al.}\cite{Mayrhofer2013}, so we will only shortly describe the concept: In a purely elastic scattering scheme, the scattering process can be described by the time-independent Schr\"{o}dinger equation. The atom-surface interaction potential is considered to be statically corrugated and periodic. Due to the surface periodicity, the interaction potential as well as the wave function can be Fourier expanded. Substitution into the Schr\"{o}dinger equation results in a set of coupled second order differential equations which are solved\cite{Sanz2007,Mayrhofer2013}.\\
The corrugated Morse potential (CMP) has proven to be a good approximation of the atom-surface interaction potential in the case of semimetal surfaces\cite{Mayrhofer2013,Kraus2015} and was applied for a first analysis of Bi$_2$Te$_3$(111). The CMP can be written as
\begin{equation}
 V(\mathbf{R},z) ~=~ D \left[ \frac{1}{\upsilon_0} e^{-2\chi(z-\xi(\mathbf{R}))} ~-~ 2e^{-\chi z} \right],
 \label{eq:CMP}
\end{equation}
with the position $\mathbf{R}$ parallel to the surface and $z$ perpendicular to the surface. Here $D$ is the potential well depth, $\chi$ is the stiffness parameter, and $\upsilon_0$ the surface average over $e^{2\chi \xi(\mathbf{R})}$. By optimising $\xi(\mathbf{R})$ as described above, a surface corrugation function of the system He-Bi$_2$Te$_3$(111) is obtained. Hereby $\xi(\mathbf{R})$ follows the hexagonal periodicity of the surface and can be written using a simple Fourier ansatz where only the amplitude is varied\cite{Mayrhofer2013} (see the inset in the lower panel of \autoref{fig:Diffraction}).\\
The calculated purely elastic intensities are corrected with the DW attenuation using the Debye temperature as determined later in this work before comparison with the experimental results. The comparison of the theoretically evaluated and corrected intensities with the experimental results was performed by determining the experimental peak areas (red squares in \autoref{fig:Diffraction}). The reason for using the peak areas instead of the peak heights is the need to account for  the broadening of the elastic peaks caused by the energy spread of the He beam. Furthermore, additional broadening of the diffraction peaks is caused by the geometry of the apparatus, defects and the domain size effects of the crystal surface.\\
Using a well depth $D = 5.8~\mathrm{meV}$ and stiffness $\chi = 0.88~\mathrm{\AA}^{-1}$\cite{Tamtogl2016} we obtain the best fit to the experimental data for a peak to peak corrugation of $0.39~\mathrm{\AA}$ or 9\% in terms of the lattice constant, respectively. The results from the CC calculation illustrated by the green crosses in \autoref{fig:Diffraction} are in good agreement with the measured peak areas, except for a slight asymmetry in the angular scan along $\overline{\mathrm{\Gamma M}}$ which is probably due to a not totally perfect alignment.\\
An electronic corrugation with a peak to peak height of 9\% of the lattice constant is considerably larger than the values found for most low-index metal surfaces\cite{Farias1998,Tamtogl2015}. The low corrugation on metal surfaces is due to the metallic charge density that smears itself out in order to lower its energy (Smoluchowski effect). This effect seems to be much reduced on the topological insulator Bi$_2$Te$_3$(111). Indeed the same has been observed for the single-component semimetal surfaces of Bi(111) and Sb(111) which yielded a similar electronic corrugation\cite{Mayrhofer2013,Kraus2015}. Since both are semimetals in the bulk with metallic surface states\cite{Hofmann2006,Sugawara2006} as well as essential components in the group of topological insulators a similar electronic corrugation for Bi$_2$Te$_3$(111) may be anticipated. Moreover, the electronic corrugation of Bi$_2$Te$_3$(111) is still smaller than the corrugation of semiconductor surfaces that have been reported so far\cite{Farias1997,Farias1998,Cardillo1981,Lambert1987,Laughlin1982}. Hence the reported value is intermediary between those of metals and semiconductors which may originate in the fact that only the surface state is metallic and the concentration of surface electrons at relatively small momentum. To our knowledge, there are no previous experimental reports on the electronic corrugation of other topological insulator surfaces and previous He atom scattering measurements\cite{Howard2013} could not provide an appropriate signal-to-noise ratio for a detailed analysis. 

\subsection{The Debye-Waller factor in scattering}
\begin{figure}
\centering
\includegraphics[width=0.48\textwidth]{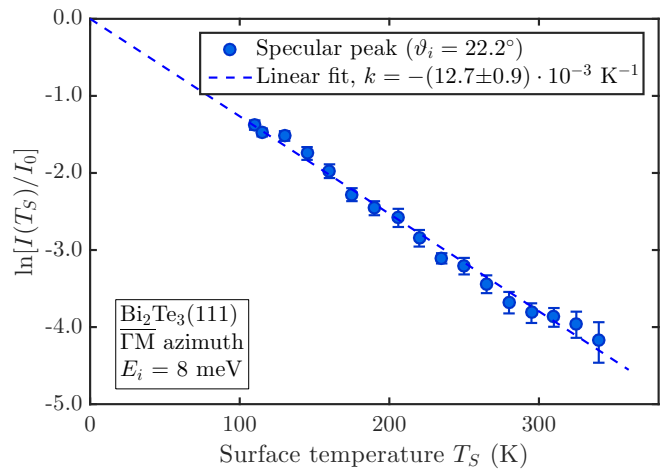}
\caption{Decay of the logarithmic specular peak intensity $\ln [I(T_S )]$ versus surface temperature $T_S$ using an incident beam energy of 8 meV for Bi$_2$Te$_3$(111)}
\label{fig:DWplot00}
\end{figure}

Upon scattering on a surface, the thermal vibrations of the surface atoms give rise to inelastic scattering events. This can be observed in the thermal attenuation of the coherent diffraction intensities without changes of the peak shape\cite{Farias1998,Tamtogl2015}. The decay of the diffraction peak intensities with increasing surface temperature $T_S$ is caused by the increasing vibrational amplitude of the surface oscillators. Its effect on the scattered intensity can be described by the DW factor, $\exp[-2W(T_{S})]$, which relates the diffraction intensity $I(T_{S})$ of a sample at temperature $T_{S}$ to the intensity $I_{0}$ for the sample at rest by\cite{Farias1998,Tamtogl2015}:  
\begin{equation}
 I(T_{S})=I_{0} \, \mathrm{e}^{-2W(T_{S})}
\label{eq:DebyeAttenuation} 
\end{equation}
Hence the thermal attenuation of the diffraction peaks provides information about the surface vibrational dynamics.\\
According to \eqref{eq:DebyeAttenuation}, the DW exponent can be determined from a plot of the natural logarithm of the intensity $\ln[I(T_S ) / I_0 ]$ versus the surface temperature $T_S$. \autoref{fig:DWplot00} shows the decay of the specular peak intensity versus the surface temperature for Bi$_2$Te$_3$(111) which gives rise  to a linear decay as expected within the Debye model. Therefore, scans of the scattered intensity versus the incident angle $\vartheta_i$ were collected while the crystal temperature was varied between 110 K and 355 K. Several of these angular scans are depicted in the inset of \autoref{fig:DWplot2} showing the decay of the first order diffraction peak for He scattered from Bi$_2$Te$_3$(111) at an incident beam energy $E_i = 8~\mbox{meV}$.\\
The observed linearity within the measured temperature range supports also the use of the high temperature approximation as applied by Manson \emph{et al.}\cite{Manson2016,Manson2016a}, which leads to \eqref{eq:DWFactor_eph}, later used in this work. Only upon a careful inspection of the data shown in \autoref{fig:DWplot00} one might anticipate a slight change of the slope at about 300 K, which may be interpreted as due to the switch-on of optical phonons at higher temperatures. However, this trend is still within the uncertainty of the measurements and we can safely employ a linear fit to the present data.

\subsection{Surface Debye temperature of Bi$_2$Te$_3$(111)}
\begin{figure}
\centering
\includegraphics[width=0.48\textwidth]{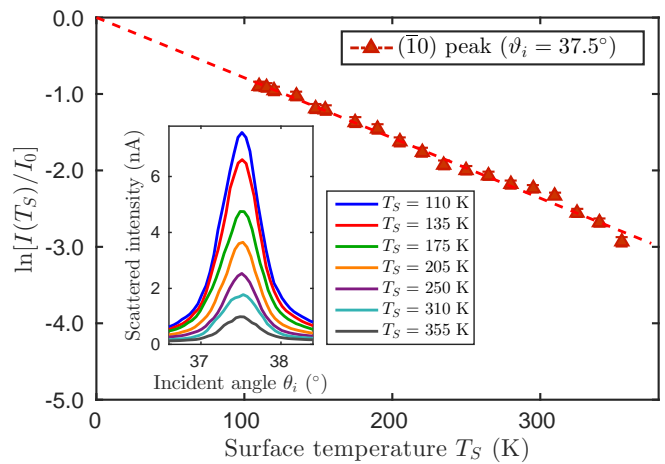}
\caption{Decay of the logarithmic peak intensity $\ln [I(T_S )]$ with increasing surface temperature $T_S$ of the first order diffraction peak along $\overline{\mathrm{\Gamma M}}$ measured at an incident beam energy of 8 meV. In the inset angular scans over the peak are depicted for a selection of different temperatures.}
\label{fig:DWplot2}
\end{figure}

We consider first the surface Debye temperature classically derived for X-ray and neutron scattering and later adapted to surface scattering. The DW exponent is given by the displacement $\mathbf{u}$ of the lattice atoms out of their equilibrium position and the momentum transfer $\mathbf{\Delta k}$ via:
\begin{equation}
 2 W(T_{S}) = \left\langle (\mathbf{u} \cdot \mathbf{\Delta k} )^2 \right\rangle_{T_{S}}
\label{eq:DWFactor_1} 
\end{equation}
where the outer brackets denote the thermal average. Since a detailed derivation of the DW factor for HAS can be found in the literature\cite{Farias1998}, it is only briefly outlined below.\\
In the case of HAS, the momentum transfer $\mathbf{\Delta k}$ can be separated into a component parallel to the surface and perpendicular to the surface (designated by subscript z): $\Delta \mathbf{k} = ( \mathbf{\Delta K}, \Delta k_{z} )$. For elastic scattering the momentum transfer parallel to the surface is given by $\lvert \mathbf{\Delta K} \rvert = \lvert \mathbf{k_i} \rvert  ( \sin (\vartheta_f ) - \sin (\vartheta_i ) )$ with the incident wave vector $\mathbf{k_i}$ and $\vartheta_i$ and $\vartheta_f$ the incident and final angle with respect to the surface normal, respectively. Within a reasonable approximation for final angles not too different from the incident angle, \eqref{eq:DWFactor_1} reduces to the $z$-component. Replacing the atom displacement with the relation for a classical harmonic oscillator together with the definition of the Debye temperature gives:
\begin{equation}
 2 W(T_{S})=\frac{3 \hbar^2\Delta k_{z}^2 T_{S}}{Mk_{B}\Uptheta_{D}^2}
\label{eq:DWFactor}
\end{equation} 
where $M$ is the mass of the surface atom, $k_B$ the Boltzmann constant and $\Uptheta_D$ the surface Debye-temperature. Since we are dealing with comparatively small parallel momentum transfers (8 meV beam) in the presented HAS experiments, equation \ref{eq:DWFactor} can be considered to be approximately correct\cite{Tamtogl2013,Tamtogl2015} and will form the basis for the following analysis.\\
However, it should be noted that \eqref{eq:DWFactor} is not generally valid for atom-surface scattering\cite{Gumhalter1996,Siber1997,Daon2012} and corrections regarding the surface phonon spectrum\cite{Levi1979} and the presence of the attractive atom-surface interaction might be necessary for other kinematic conditions. The influence of the attractive part of the atom-surface potential upon scattering can be considered by the Beeby correction\cite{Beeby1971}. The perpendicular momentum transfer $\Delta k_z$ is then replaced by\cite{Farias1998}:
\begin{equation}
\Delta k_z ^{'}  = k_i \left[ \sqrt{ \cos ^2 (\vartheta_f) + \frac{D}{E_i}} + \sqrt{ \cos ^2 (\vartheta_i) + \frac{D}{E_i}}~\right]
\label{eq:BeebyCorr}
\end{equation}
which assumes an attractive part of the potential with a spatially uniform well of depth $D$. In the case of the specular geometry, $\vartheta_i = \vartheta_f$ holds and the DW exponent \eqref{eq:DWFactor} together with the Beeby correction further simplifies to:
\begin{equation}
2 W(T_{S}) = \frac{24m \left[ E_i \cos ^2 (\vartheta_i) + D\right] T_{S}}{M k_{B} \Uptheta_{D}^2}
\label{eq:DebyeWallerSpec} 
\end{equation} 
where $m$ is the impinging particle mass and the momentum has been replaced by the incident beam energy $E_i$ using $k_i^2=2m  E_i / \hbar^2$.\\
The surface Debye temperature can then be calculated using the slope of a linear fit of $\ln[I(T_S ) / I_0 ]$ versus the surface temperature $T_S$.
We can again make use of \autoref{fig:DWplot00} which shows the decay of the natural logarithm of the specular peak intensity versus the surface temperature for Bi$_2$Te$_3$(111). \autoref{eq:DebyeWallerSpec}, together with \eqref{eq:DebyeAttenuation} is used to determine the surface Debye temperature ($\Uptheta_D$) from the experimental data. We have used a value of $D = 5.8~\mbox{meV}$ for the potential well depth in the present analysis as already mentioned in \autoref{sec:elastic}. (The sensitivity of the Debye temperature $\Uptheta_D$ to changing $D$ by 1~meV is relatively small and falls within the experimental uncertainty of $\Uptheta_D$).\\
One must also assume a value for the mass $M$ which is typically the mass of the crystal atoms since the surface Debye temperature is related to the motion of the ion cores. There is some ambiguity connected with the mass of the surface scatterer, since this concept was adapted from other scattering techniques and He atoms are scattered by the electron density rather the ion cores as described above\cite{Tamtogl2013a,Hayes2007,Taleb2016}. Due to the e-ph coupling, the surface Debye temperature is associated with charge density oscillations that are induced by vibrations of the ion cores which will be disussed in \autoref{sec:EPhDW}. Hence, strictly speaking, when writing the DW exponent in the usual way as a quantity directly proportional to the surface atom mean-square displacements, one needs to consider an effective surface Debye temperature $\Uptheta_D^{*} $ and a surface atom effective mass $M^{*}$, which includes contributions arising from the e-ph interaction. Nevertheless, these simple equations have proven to serve as a good approximation in the case of HAS\cite{Farias1998,Politano2011,Becker2010} and using the mass of a single surface atom is a reasonable choice in most cases\cite{Shichibe2015,Politano2011,Tamtogl2015}.\\
Using the best-fit result of \autoref{fig:DWplot00}, a surface Debye temperature $\Uptheta_D = (81\pm6)~\mbox{K}$ is obtained for the specular geometry. Here we have set $M$ in \eqref{eq:DebyeWallerSpec} equal to the mass of a single Te atom.\\ 
In order to confirm the consistency of our measurements we apply the same analysis to the first order diffraction peak measured along the $\overline{\mathrm{\Gamma M}}$ azimuth. A plot of $\ln [I (T_S )]$ versus the surface temperature $T_S$ for the first order diffraction peak is depicted in figure \ref{fig:DWplot2}. Using the slope of the linear fit, the Debye temperature can be calculated with \eqref{eq:DWFactor}. Since the mirror condition $\vartheta_i = \vartheta_f$ no longer holds, the perpendicular momentum transfer is calculated using \eqref{eq:BeebyCorr} with the same parameters for the well depth as before. The surface Debye temperature from the analysis of the first order diffraction peak intensities is $\Uptheta_D = (97\pm8)~\mbox{K}$. The value is slightly larger but still in reasonable agreement with the Debye temperature determined from the specular peak.\\
Both values are significantly reduced with respect to the bulk value which has been determined as $\approx 145~\mathrm{K}$\cite{Walker1960}. This is in good agreement with simple theoretical approximations that estimate a reduction by a factor of $1/\sqrt{2}$ with respect to the bulk value\cite{Delft1991}. Furthermore, the surface Debye temperature of Bi$_2$Te$_3$(111) is very similar to the surface Debye temperature of Bi(111)\cite{Mayrhofer2012,Hofmann2006}.

\subsection{Electron-phonon coupling from the Debye-Waller factor}
\label{sec:EPhDW}
\begin{figure*}[htb]
\centering
\includegraphics[width=0.9\textwidth]{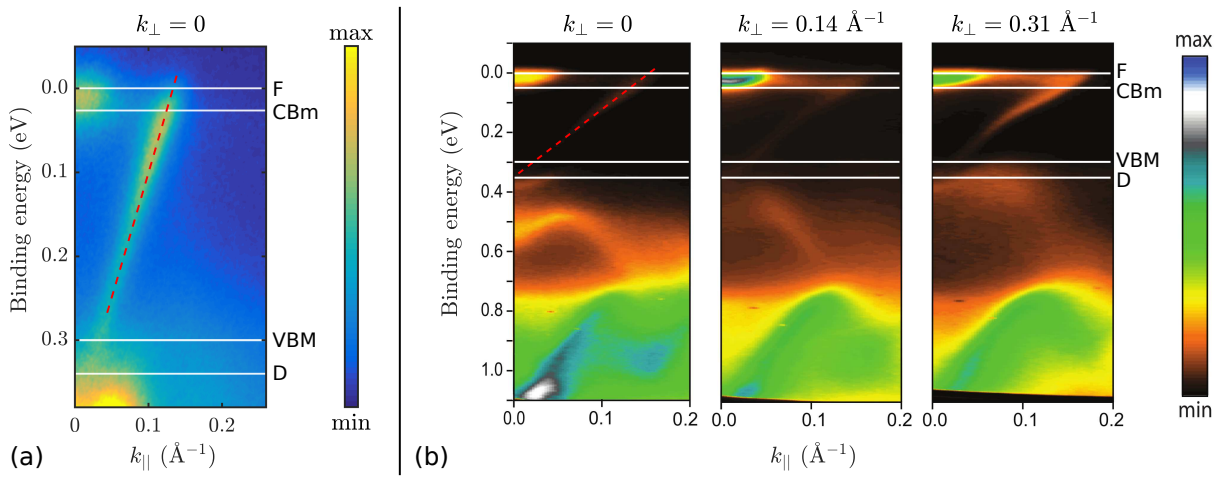}
\caption{(Colour online) Photoemission intensity for Bi$_2$Te$_3$(111) along $\overline{\mathrm{\Gamma M}}$ of (a) the sample of the present HAS study and (b) the sample measured by Michiardi \emph{et al.} (Adapted figure with permission from \cite{Michiardi2014}, Copyrighted by the American Physical Society) at three different values of the normal momentum transfer (with the zero conventionally set at the $\overline{\Gamma}$-point). The positions of the Fermi level (F), the conduction band minimum (CBm), the valence band maximum (VBM) and the Dirac point (D) are indicated by the horizontal lines. The dashed red lines show the quasi-linear dispersion of the surface Dirac states.}
\label{fig:ARPES}
\end{figure*}

As already mentioned above, in the case of HAS the He atoms are not scattered by the vibrating ion cores. Here, inelastic scattering corresponds to a scattering by phonon-induced charge density oscillations\cite{Benedek2014,Tamtogl2013a,Senet2002}. These charge density oscillations are related to the surface Debye temperature via the e-ph coupling.\\
Upon scattering of a He atom, the probability of creating (or annihilating) a phonon of frequency  $\omega_{ \mathbf{Q} , \nu }$ $\{ \mathbf{Q}, \nu \}$ with parallel wavevector $\mathbf{Q}$ and branch index $\nu$, is proportional to the corresponding mode-specific e-ph coupling constant $\lambda_{ \mathbf{Q},\nu }$\cite{Manson2016,Benedek2014}. The DW exponent is then expressed as a sum over all contributing phonon modes where each phonon mode ${\mathbf{Q},\nu}$ is weighted by the mode-specific $\lambda_{\mathbf{Q},\nu}$. For the simplest case, the specular diffraction peak, Manson \textit{et al.} derived the following DW exponent:\cite{Manson2016,Manson2016a} 
\begin{equation}
 2 W (\mathbf{k_f},\mathbf{k_i}) \cong 4 \, \mathcal{N} (E_F ) \: \frac{m}{m_e^{*}} \frac{E_{iz}}{\phi} \: \lambda \: k_B \: T_S
\label{eq:DWFactor_eph} 
\end{equation}
for scattering of an incoming He atom with the wavevector $\mathbf{k_i}$ into a final state of wavevector $\mathbf{k_f}$. Here $\mathcal{N} (E_F )$ is the density of electron states per unit surface cell at the Fermi level and $\phi$ is the work function. $m$ is the mass of the $^3$He atom, $m_e^{*}$ is the electron effective mass and the mass enhancement factor $\lambda$ expresses now the total e-ph coupling strength. $E_{iz}$ is the beam energy $E_i$ normal to the surface ($E_{iz}=E_i \cos \vartheta_i$) and $k_B$ is the Boltzmann constant. As explained above, the normal incident energy $E_{iz}$ needs to include the Beeby correction, i.e., to be replaced by $E_{iz}^{'} = E_{iz} + D$.\\
As appears from \eqref{eq:DebyeAttenuation} and \eqref{eq:DWFactor_eph}, the e-ph coupling constant $\lambda$ can be directly obtained from the temperature dependence of the HAS specular intensity $I$ using the slope determined from the DW plot (\autoref{fig:DWplot00}, $ - \Delta \ln (I / I_0 )  / \Delta T_S = \ln (I_0 / I ) / T_S $):
\begin{equation}
\lambda_{HAS} \cong \frac{1}{ 4 \, \mathcal{N} (E_F )} \frac{m_e^{*}}{m} \frac{\phi}{k_B \: E_{iz}} \frac{ \ln (I_0 / I ) }{T_S}.
\label{eq:DW_lambda1} 
\end{equation}
This expression has been derived in \cite{Manson2016,Manson2016a} for conducting surfaces and shown to provide for the surfaces of ordinary metals and semimetals values of $\lambda$ in agreement with those known from other sources.\\
When applied to the surface of a \emph{degenerate} semiconductor with the Fermi level within the conduction band (as in the present case according to \cite{Suh2014,Michiardi2014,Pettes2013}) one has to consider that the phonon-induced modulation of the surface charge density only involves electrons near the surface within the Thomas-Fermi screening length (TFSL), $k_0^{-1} = ( 2 \epsilon E_F / 3 e^2 n_0 )^{1 / 2}$, where $\epsilon = \epsilon_r \epsilon_0$, is the static dielectric constant, $e$ the electron charge and $n_0$ the carrier density in the conduction band\cite{Ashcroft}. $\mathcal{N} (E_F )$ is then written as $\mathcal{N} (E_F ) = 3 n_0 A_{c} / 2 E_F k_0 $ where $A_{c}$ is the surface unit cell area, and the Fermi energy $E_F$ is referred to the conduction band minimum. With the substitutions $n_0 = k_{F \parallel}^2 k_{F \perp } / 3 \pi^2$, valid for an anisotropic 3D free electron gas, where $k_{F \parallel}$ and $k_{F \perp }$ are the Fermi wavevectors of the conduction electrons parallel and normal to the surface, and $E_F = \hbar^2 k_{F \parallel}^2 / 2 m_e^{*}$ it is found:
\begin{equation}
 \mathcal{N} (E_F ) = \frac{ m_e^{*} A_c }{ \pi \hbar^2 } \frac{ k_{F \perp } }{ \pi k_0 }
 \label{eq:ElDOS}
\end{equation}
The first factor is easily recognized as the 2DEG Fermi-level density of states referred to the surface unit cell. As to the second factor, since in n-Bi$_2$Te$_3$ and similar materials $\epsilon$ is very large (e.g., $\epsilon_r = 290$ in n-Bi$_2$Te$_3$(111)\cite{Richter1977}), it is in general $k_{F \perp } / \pi k_0 \gg 1$. Thus the major contribution to $\mathcal{N} (E_F )$ effective in the total surface e-ph interaction comes from the surface-projected DOS of the conduction electron states above the conduction band minimum (CBm, \autoref{fig:ARPES}) rather than from the Dirac states. As appears from ARPES data on the 3D material\cite{Michiardi2014} the conduction band in the normal direction is rather flat and $k_{F \perp }$ is hard to extract.\\
Thus $\lambda_{HAS}$ is more conveniently expressed in terms of $k_0$, $k_{F \parallel }$ and $n_0$, the latter being known from transport measurements:
\begin{equation}
\begin{split}
\lambda_{HAS} & \cong \alpha \: \frac{ k_{ F \parallel }^2 \: k_0 }{ 6 n_0 } = \alpha \: \frac{ m_e^{*}}{\hbar^2} \sqrt{ \frac{ E_F \: e^2 }{ 6 \: \epsilon \: n_0 } }, \\
 \alpha & \equiv   \frac{ \phi \: \ln (I_0 / I ) } { A_c \: k_{iz}^2 \: k_B  \: T_S}
\label{eq:DW_lambda2} 
\end{split}
\end{equation}
where $k_{iz} = \sqrt{2 m E_{iz}^{'}} / \hbar$ is the normal component of the He incident wavevector including the Beeby correction. Using the slope determined from the linear fit in \autoref{fig:DWplot00}, together with $\phi=4.9~\mathrm{eV}$\cite{Suh2014} and $A_{c} = 16.5~\mathrm{\AA}^2$ (using the lattice constant $a$ determined from the diffraction data in \autoref{sec:elastic}) gives $\alpha = 2.38$. With $E_F = 0.025~\mathrm{eV}$, as suggested by ARPES data on the present sample (\autoref{fig:ARPES}(a)) and reported by Suh \emph{et al.} for unirradiated samples\cite{Suh2014} and $m_e^{*} = 0.07 m_e$(from Huang \emph{et al.}\cite{Huang2008}), \eqref{eq:DW_lambda2} gives:
\begin{equation}
\lambda_{HAS} \cong 0.11 / \sqrt{ n_0 }
\label{eq:HAS_lambda} 
\end{equation}
with $n_0$ given in units of $10^{20}~\mathrm{cm}^{-3}$. As discussed by Suh \emph{et al.}\cite{Suh2014}, unirradiated n-Bi$_2$Te$_3$ thick films show a conduction electron density of about $7 \cdot 10^{18}~\mathrm{cm}^{-3}$. Note that the ARPES measurements by Michiardi \emph{et al.}\cite{Michiardi2014} of the 3D band structure for a similar sample, reproduced in \autoref{fig:ARPES}(b) for the $\overline{\Gamma \mathrm{M}}$ direction, indicate a Fermi level above the CBm of $\approx 0.05~\mathrm{eV}$, which would give $\lambda_{HAS} \cong 0.15 / \sqrt{ n_0 }$.\\
The carrier concentration from Hall measurements in the present sample is $6.44\cdot 10^{18}~\mathrm{cm}^ {-3}$, in agreement with the data of Suh \emph{et al}\cite{Suh2014}, for  unirradiated n-Bi$_2$Te$_3$ thick films ($7\cdot 10^{18}~\mathrm{cm}^ {-3}$). This would give $\lambda_{HAS} \cong 0.42$ for $E_F = 0.025~\mathrm{eV}$ (our sample) or $\lambda_{HAS} \cong 0.57$ for $E_F = 0.05~\mathrm{eV}$. However, as shown in the work by Suh \emph{et al.}\cite{Suh2014}, the carrier concentration rapidly increases with decreasing thickness of the samples, due to the increasing weight of the surface accumulation layer in the region within the TFSL where the downward band bending occurs\cite{Pettes2013}. As explained above, it is the carrier density near the surface, the one within the TFSL, which enters \eqref{eq:HAS_lambda}. Hence a carrier density in the range of  $10^{20}~\mathrm{cm}^ {-3}$, corresponding to $\lambda_{HAS} \cong 0.1$ or less, appears to be more appropriate.\\
Indeed recent \emph{ab-initio} calculations by Huang\cite{Huang2012} for 3 QLs found $\lambda = 0.05$ in agreement with Heid \emph{et al.}\cite{Heid2016} which obtain for a calculation of 5 QLs as a function of the Fermi energy the same value when $E_F$ is $0.35~\mathrm{eV}$ above the Dirac point (as in \autoref{fig:ARPES}). This value of $\lambda$ would require an average density of $n \approx 4\cdot 10^{20}~\mathrm{cm}^ {-3}$ for the 2DEG in the surface band bending region.\\
\begin{table}\centering
\caption{Comparison of the electron-phonon coupling constants $\lambda$ for Bi$_2$Te$_3$(111) obtained from both experimental methods and calculations. The mode-specific $\lambda$ for the low-lying surface optical mode is much larger than the average $\lambda$ as listed in the second part of the table. We obtain a lambda of $0.11 / \sqrt{n_0}$ (with $n_0$ in $10^{20}~\mathrm{cm}^{-3}$) which corresponds to a range of $0.04-0.11$ for a charge carrier density of the order of $10^{20}~\mathrm{cm}^{-3}$.}
\begin{tabular}{ l l l lS[table-format=1.2] }
\hline
\multicolumn{1}{l }{Sample} & \multicolumn{1}{l }{Methods} &\multicolumn{1}{l }{Ref.$~$ } & \multicolumn{1}{c }{$\lambda$} \\ 
\hline
\hline
\rule{0pt}{1\normalbaselineskip}cleaved & ARPES\footnote{With the contribution from both the e-ph coupling and the electron-spin-plasmon coupling\cite{Kondo2013}} &\cite{Kondo2013} & 3 \\ 
theory semi-$\infty$  & PCM$+$LRT\footnote{Mode-selected e-ph coupling for the low-lying surface optical mode using a pseudo-charge model (PCM) for lattice dynamics and linear response theory} &\cite{Howard2014} & 2\hfill \\ 
cleaved  & HAS\footnote{Mode-selected e-ph coupling for the surface optical mode originating at 5.8 meV at the $\overline{\Gamma}$-point} & \cite{Howard2013} & 1.44 \\ 
\hline
\rule{0pt}{1\normalbaselineskip}cleaved n-type & ARPES & \cite{Chen2013} & 0.19 \\ 
theory semi-$\infty$ & IEC\footnote{Isotropic elastic continuum (IEC) model for lattice dynamics} & \cite{Giraud2011} & 0.13 \\ 
cleaved n-type & HAS\footnote{Present work with the carrier density $n_0$ in $10^{20}~\mathrm{cm}^{-3}$} &  & $0.11 / \sqrt{n_0}$\\
theory 3 QL & DFPT $+$ SOC\footnote{Density functional perturbation theory (DFPT) with spin-orbit coupling (SOC)} & \cite{Huang2012} & 0.05 \\ 
theory 5 QL  & DFPT $+$ SOC\footnote{For a Fermi level 0.35 eV above the Dirac point} & \cite{Heid2016} & 0.05 \\ 
cleaved p-type & ARPES & \cite{Chen2013}&  0.0 \\ 
\hline
\end{tabular}
\label{tab:LambdaTable}
\end{table}
A listing of the e-ph coupling on Bi$_2$Te$_3$(111) obtained from both experimental methods and calculations can be found in \autoref{tab:LambdaTable}. It is immediately apparent that there exists a large variation in the e-ph coupling constants determined for Bi$_2$Te$_3$(111), in particular between a mode-selected e-ph coupling associated with some prominent Kohn anomaly, as obtained from inelastic HAS\cite{Howard2013,Howard2014} and the value of $\lambda$ averaged over the whole phonon spectrum as derived from ARPES measurements\cite{Chen2013}. One reason for this discrepancy is that the rather large values for a specific phonon mode\cite{Howard2013,Howard2014} cannot be compared with ARPES measurements where one typically integrates over all phonon modes. Furthermore, the exceptionally large value determined by Kondo \textit{et al.}\cite{Kondo2013} is dominated by the contribution from a theoretically proposed spin plasmon\cite{Kondo2013}.\\
The great variability of $\lambda$ values reported for Bi$_2$Te$_3$ and similar materials largely depends on their complex electronic structure and the fact that different e-ph dependent properties actually enucleate different aspects of the e-ph interaction. In insulators and non-degenerate semiconductors the e-ph interaction is related to the polarization of valence electrons due to phonon displacements and related electric fields, and essentially involves virtual electronic excitations across the gap. In metals and degenerate semiconductors the phonon-induced electronic transitions between states near the Fermi level play the major role. In conducting chalcogenides with large anion polarizabilities both aspects of e-ph interaction can be equally important. In narrow-gap topological insulators with only surface Dirac states crossing the Fermi level, the role of valence electrons in $\lambda$ is important; even more so if they are degenerate semiconductors with the Fermi level lifted above the CBm as a consequence of doping or surface defects, as in the present case.\\
In all these complex cases better information would be obtained from electron state-selected e-ph interaction, as from ARPES or recent Heid \emph{et al.} \emph{ab-initio} studies\cite{Heid2016}, or phonon mode-selected e-ph coupling constants, as from inelastic HAS\cite{Benedek2014}. However, both the mode-selected ($\lambda_{\mathbf{Q},\nu}$) and the average ($\lambda$) e-ph constants derived from HAS contain the contribution of all electronic transitions producing a modulation of the surface charge density. Even in an insulating surface such as Xe(111) a quantum-sonar effect is observed with HAS, which is a manifestation of e-ph interaction due to the large polarizability of Xe atoms\cite{Campi2015}. Thus in the general case of different contributions to the e-ph interaction, the temperature dependence of the HAS DW exponent provides a complete information on $\lambda$ at the surface by directly measuring the electron charge-density oscillations induced by the entire surface-projected phonon spectrum.\\
Hence, combined with ARPES data, the new approach based on the DW attenuation in He scattering enables us to extract the correctly averaged $\lambda$ for a given surface\cite{Manson2016}. Furthermore, the measurement of the thermal attenuation is a robust and reliable experimental method to extract the e-ph coupling. In contrast, the determination of the e-ph coupling directly from ARPES, requires high resolution measurements due to the small momentum space and energy window relevant for this process\cite{Chen2013} and it remains difficult to exclude effects from bulk states\cite{Sobota2014}.\\
When considering the average $\lambda$ of  Bi$_2$Te$_3$(111), our measurements suggest a rather modest e-ph coupling in agreement with other studies (\autoref{tab:LambdaTable}). While it is difficult to directly relate superconducting states with the e-ph coupling of a surface, the fact that Bi$_2$Te$_3$ does not exhibit a superconducting phase under ``normal'' conditions\cite{Matsubayashi2014,Le2014} supports our results. Hence our finding of a modest $\lambda$ is not surprising even though some individual phonons may provide a much larger e-ph interaction as reported by Howard \textit{et al.}\cite{Howard2013}.\\

\section{Summary and Conclusion}
We have studied the topological insulator Bi$_2$Te$_3$(111) by means of helium atom scattering. The exact close-coupling method was applied to determine the electronic surface corrugation from the diffraction peak intensities. Hereby Bi$_2$Te$_3$(111) exhibits an electron density corrugation with a peak to peak height of $0.39~\mbox{\AA}$ (9\% of the surface lattice constant) upon scattering of $^3$He with a beam energy of 8 meV. The obtained electronic corrugation is intermediary between those of metal and semiconductor surfaces as well as comparable to studies of semimetal surfaces. The thermal attenuation in the diffraction of He from Bi$_2$Te$_3$(111) was studied in a temperature range between $110~\mbox{K}$ and $335~\mbox{K}$. The system shows a typical Debye-Waller behaviour with a surface Debye temperature of $\Uptheta_D = (81\pm6)~\mbox{K}$.\\
By adapting a recently developed quantum-theoretical derivation of the He scattering probabilities to the case of a degenerate semiconductor, we are able to extract the e-ph coupling strength from temperature dependent He atom scattering measurements. The Debye-Waller attenuation, which is directly related to the e-ph coupling $\lambda$, allows to extract a correctly averaged $\lambda$ for a given surface. Hence we obtain a $\lambda$ in the range of $0.04-0.11$ for Bi$_2$Te$_3$(111) for a charge carrier density of the order of $10^{20}~\mathrm{cm}^{-3}$. This relatively modest value of $\lambda$ is not surprising even though some individual phonons may provide a larger e-ph interaction. Our results of the e-ph coupling are consistent with other theoretical and experimental results and suggest that previous inconsistencies were found when relating mode-specific values with an average $\lambda$.\\
With respect to potential applications, the obtained weak electron-phonon coupling indicates that the Bi$_2$Te$_3$ system may be useful in achieving high electron mobilities. In particular the growth of defect-free ultrathin films offers a route towards fast devices based on any small-gap semiconductors such as Bi$_2$Te$_3$ due to the high electron mobilities (small $m_e^{*}$) and a long mean-free path (small e-ph interaction).

\section{Acknowledgement}
The authors would like to thank J. R. Manson for many helpful discussions. One of us (A.T.) acknowledges financial support provided by the FWF (Austrian Science Fund) within the project J3479-N20. The authors are grateful for financial support by the Blavatnik Foundation, the Aarhus University Research Foundation, VILLUM FONDEN via the Centre of Excellence for Dirac Materials (Grant No. 11744) and the SPP1666 of the DFG (Grant No. HO 5150/1-1). M.B., E.M.J.H. and B.B.I. acknowledge financial support from the Center of Materials Crystallography (CMC) and the Danish National Research Foundation (DNRF93).
 
\bibliography{literature}

\end{document}